\documentclass[showpacs,twocolumn]{revtex4}
\usepackage{amsmath}
\usepackage{amssymb}
\usepackage{bm}
\usepackage{dcolumn}
\usepackage{graphics}
\usepackage{graphicx}
\usepackage{epsfig}
\usepackage{amssymb}

\begin{document}

 \title{Exponential suppression of interlayer conductivity in
very anisotropic quasi-two-dimensional compounds in high magnetic
field}
 \author{P. D. Grigoriev}
 \affiliation{L. D. Landau Institute for Theoretical Physics, Chernogolovka, 142432, Russia}
 \altaffiliation{Temporarily at Laboratoire de Physique des Solides, Universite Paris-Sud 11, 91405, Orsay, France}
 \email{grigorev@itp.ac.ru}

\date{\today }

\begin{abstract}
It is shown that in rather strong magnetic field the interlayer
electron conductivity is exponentially damped by the Coulomb
barrier arising from the formation of polaron around each
localized electron state. The theoretical model is developed to
describe this effect, and the calculation of the temperature and
field dependence of interlayer magnetoresistance is performed. The
results obtained agree well with the experimental data in
GaAs/AlGaAs heterostructures and in strongly anisotropic organic
metals. The proposed theory allows to use the experiments on
interlayer magnetoresistance to investigate the electron states,
localized by magnetic field and disorder.
\end{abstract}


 \maketitle

\section{Introduction}

Layered materials is a wide class of compounds of very strong
scientific and technological impact. These materials are composed
of a stack of two-dimensional (2D) conducting layers with weak
interlayer electron coupling, determined by the interlayer
transfer integral $t_{z}$. Such layered structure may be of
natural molecular origin, as, e.g., in cuprates, pnictides,
organic metals or graphite, or of artificial origin, as in
heterostructures or intercalated graphites. The small value of
$t_{z}$, which is much smaller than the in-plane Fermi energy
$E_{F}$, provides these compounds with highly anisotropic
electronic properties.

Magnetoresistance is traditionally used to explore the internal electron
structure of metals.\cite{Ziman,Abrik,Shoenberg} Magnetoresistance in
layered compounds has many special features. These are beats of magnetic
quantum oscillations (MQO),\cite{Shoenberg} angular magnetoresistance
oscillations (AMRO),\cite{MK1988,Yam,Yagi} slow oscillations of
magnetoresistance\cite{SO,Shub}, the phase shift of beats of MQO,\cite%
{PhSh,Shub} etc. \ The angular dependence of magnetoresistance is widely
used to investigate the electronic structure of various layered compounds:
organic metals (see, e.g., Refs. \cite%
{KartPeschReview,MarkReview2004,OMRev,MQORev} for reviews), cuprate
high-temperature superconductors,\cite%
{HusseyNature2003,AbdelNature2006,AbdelPRL2007AMRO,McKenzie2007}
heterostructures\cite{Kuraguchi2003} etc. The AMRO are well understood\cite%
{Yam,Yagi} in the framework of the standard semiclassical 3D theory, based
on the Boltzmann transport equation, which gives the Shockley-Chambers
formula\cite{Ziman} for the angular dependence of magnetoresistance. The
recent theoretical works\cite{HusseyNature2003,Bergemann,PDG2010} on AMRO
only provide the calculations based on this formula, which are useful for
its practical application to extract the Fermi-surface parameters from the
experimental data.

There is one important feature of the angular dependence of
interlayer magnetoresistance, observed in various strongly
anisotropic layered compounds, which cannot be described in the
framework of the standard 3D theory. This is the so-called
coherent-incoherent transition of interlayer magnetoresistance,
induced by disorder and by magnetic field component $B_{z}$
perpendicular to the conducting layers.\cite%
{WosnitzaPRL2001,Wosnitza2002,MarkPRL2006,Incoh2009} This transition occurs
in layered compounds with extremely high quasi-2D anisotropy, namely, when
the interlayer transfer integral $t_{z}$ is less than the cyclotron energy $%
\hbar \omega _{c}=\hbar eB_{z}/m^{\ast }c$ and than the Landau level (LL)
broadening $\Gamma _{0}$ due to impurities. This coherent-incoherent
transition or crossover is sometimes accompanied by the metal-insulator
transition\cite{Zuo1999,WosnitzaPRL2001} and is always characterized by a
strong growth of interlayer magnetoresistance $R_{zz}$ with the increase of $%
B_{z}$,\cite%
{WosnitzaPRL2001,Wosnitza2002,MarkPRL2006,Incoh2009,Zuo1999,Nam2001} while
the component $B_{\parallel }$ of magnetic field parallel to the conducting
layers only slightly affects the interlayer conductivity at weak magnetic
field.\cite{MarkPRL2006,Incoh2009} It is important that $R_{zz}\left(
B_{z}\right) $ grows not only in the maxima of MQO of magnetoresistance,
when the chemical potential is situated in the gap between LLs, but also in
the minima of magnetoresistance, when the chemical potential is exactly on
the LLs and when there is no any gap of electron states at the Fermi level.
The 2D formula\cite{ChampelMineev} for the Shubnikov-de Haas effect, based
on the standard 3D theory of interlayer magnetoresistance and applied to the
extremely anisotropic quasi-2D limit, does not describe this effect.
Moreover, the first direct comparison between the magnetoresistance in the
coherent 3D and the so-called "weakly incoherent" regimes did not reveal any
considerable differences.\cite{MosesMcKenzie1999} Recently it was shown,\cite%
{PDGPRB2011,PDGWIJETP2011,PDGFNTWI2011} that although the initial model in
Ref. \cite{MosesMcKenzie1999} is valid for the weakly incoherent limit, the
calculation of the electron Green's functions in disordered 2D layer in Ref.
\cite{MosesMcKenzie1999} is incorrect, because the impurity scattering is
incorrectly treated via a constant imaginary part of the electron
self-energy similar to 3D case. More rigorous calculations show (see, e.g.
in Ref. \cite{Ando}) that in strong magnetic field, in addition to the
oscillations due to LL quantization, the imaginary part of the electron self
energy monotonically increases with magnetic field $\propto \sqrt{B_{z}}$.%
\cite{Ando} This monotonic increase leads\cite%
{PDGPRB2011,PDGWIJETP2011,PDGFNTWI2011} to the similar increase of the
Dingle temperature and of the monotonic part of magnetoresistance $\propto
\sqrt{B_{z}}$, which also changes the angular dependence of
magnetoresistance. This improvement\cite{PDGPRB2011,PDGWIJETP2011} of the
theory can explain some experimental observations\cite{Kang} of the
monotonic growth of interlayer magnetoresistance as function of $B_{z}$ when
$\Gamma _{0},t_{z}\lesssim \hbar \omega _{c}$. However, in many compounds
the much stronger than $\propto \sqrt{B_{z}}$ growth of interlayer
magnetoresistance has been observed.\cite%
{WosnitzaPRL2001,Wosnitza2002,MarkPRL2006,Incoh2009,Zuo1999,Nam2001} To
explain this rapid growth of $R_{zz}\left( B_{z}\right) $, the
variable-range hopping mechanism of interlayer electron transport has been
proposed,\cite{Gvozd2007} similar to the in-layer transport in the regime of
quantum Hall effect (QHE). Although a reasonable agreement with experiment
has been achieved in Ref. \cite{Gvozd2007} basing on the
semi-phenomenological consideration, the underlying theoretical model is
still unclear. First, the actual electron interlayer hopping is not
variable-range, as in Ref. \cite{Gvozd2007}, but always to the adjacent
layer, i.e. on the interlayer distance $d$. Moreover, the AMRO still exist
in this limit, which means that during the interlayer hopping the in-plane
electron wave function is conserved, i.e. the interlayer tunnelling term in
the Hamiltonian conserves the in-plane electron momentum. This Hamiltonian
is given by Eqs. (\ref{H})-(\ref{Hi}) below. Without electron-electron (e-e)
and electron-phonon (e-ph) interaction it has been studied in Refs. \cite%
{PDGPRB2011,PDGWIJETP2011,PDGFNTWI2011} (see Eqs. (8)-(10) of Ref. \cite%
{PDGPRB2011}). However, the comparison with some experiments shows
that the results of Refs.
\cite{PDGPRB2011,PDGWIJETP2011,PDGFNTWI2011} are applicable only
in the intermediate magnetic fields, when $\Gamma _{0},t_{z}<\hbar
\omega _{c}$, but the magnetic field is still too weak to induce
the metal-insulator transition.

The aim of the present paper is to generalize the model of Ref. \cite%
{PDGPRB2011} to describe the strong increase of interlayer magnetoresistance
and the metal-insulator transition in very high magnetic field.

\section{The model}

The quasi-2D layered electron system with disorder and interaction is
described by the following Hamiltonian, which contains four main terms:
\begin{equation}
\hat{H}=\hat{H}_{0}+\hat{H}_{t}+\hat{H}_{I}+\hat{H}_{int}.  \label{H}
\end{equation}%
The first term $\hat{H}_{0}$ is the Hamiltonian of the  noninteracting 2D
electron gas in magnetic field summed over all layers. The second term in
Eq. (\ref{H}) gives the coherent electron tunnelling between two adjacent
layers:
\begin{equation}
\hat{H}_{t}=2t_{z}\sum_{j}\int d^{2}\boldsymbol{r}[\Psi _{j}^{\dagger }(%
\boldsymbol{r})\Psi _{j-1}(\boldsymbol{r})+\Psi _{j-1}^{\dagger }(%
\boldsymbol{r})\Psi _{j}(\boldsymbol{r})],  \label{Ht}
\end{equation}%
where $\Psi _{j}(\boldsymbol{r})$ and $\Psi _{j}^{\dagger }(\boldsymbol{r})$%
\ are the creation (annihilation) operators of an electron on the layer $j$
at the point $\boldsymbol{r}$. This interlayer tunnelling Hamiltonian is
called "coherent" because it conserves the in-layer coordinate dependence of
the electron wave function (in other words, it conserves the in-plane
electron momentum) after the interlayer tunnelling. The third term\
\begin{equation}
\hat{H}_{I}=\sum_{i}\int d^{3}\mathbf{r}V_{i}\left( \mathbf{r}\right) \Psi
^{\dagger }(\boldsymbol{r})\Psi (\boldsymbol{r})  \label{Hi}
\end{equation}%
gives the electron interaction with impurity potential. The last
term $\hat{H}_{int}$ describes the electron-electron and
electron-phonon interaction. The presence of this term differs our
model from that in Refs.
\cite{PDGPRB2011,PDGWIJETP2011,PDGFNTWI2011}. Even without $\hat{
H}_{int}$ and for the case of point-like impurity potential, the
exact solution of this Hamiltonian is not achievable. After
inclusion of the interaction term $\hat{H}_{int}$, we can only
catch the main physical effects in this system to describe the
interlayer electron transport.

In the limit, $t_{z}\ll \Gamma _{0},\hbar \omega _{c}$, the
interlayer hopping $t_{z}$ can be considered as a small
perturbation for the uncoupled stack of 2D metallic layers. The 2D
disordered electron system in magnetic field has been extensively
studied in connection to the quantum Hall effect. With the
increase of magnetic field the crossover from the diffusive
dynamics and classical Shubnikov-de Haas effect to the in-plane
localization of electron states takes place. In Ref.
\cite{Fogler1997} this crossover is described in detail, and the
value of the crossover field $B_{c}$ is estimated for the various
types of impurity potential. Although for point-like impurity
potential the crossover field $B_{c}$ was not calculated in Ref.
\cite{Fogler1997}, the general criteria that $B_{c}$ corresponds
to field when the nearest Landau levels (LLs) become separated,
i.e. when $\hbar \omega _{c}\gtrsim \Gamma _{0}$, can be used.
During this crossover the localization length $\xi $ of the 2D
electron states decreases exponentially
from infinity to the value\cite%
{Fogler1997,Fogler1998,Raikh1995,HuckesteinRMP1995,AmadoPRL2011,Furlan1998,Lewis2001,Hohls2001,Hohls2002,Comment1}
\begin{equation}
\xi \approx R_{c},  \label{xi1}
\end{equation}%
where the classical cyclotron (Larmor) radius%
\begin{equation}
R_{c}=\hbar k_{F}c/eB_{z}=k_{F}l_{H}^{2}=2\nu /k_{F}, \label{Rc}
\end{equation}%
$k_{F}$ is the in-plane Fermi momentum, $l_{H}=\sqrt{\hbar c/eB_{z}}$ is the
magnetic length, and the LL filling factor $\nu = F/B_{z}$, where $%
F=k_{F}^{2}c\hbar /2e$ is the MQO frequency. In the quantum limit
$\nu \leq 1$ the localization length $\xi \approx l_{H}$.

In many layered materials the in-plane electron dispersion is anisotropic,
and the in-plane FS has the shape of elongated ellipse with semi-axes given
by the wave vectors $k_{F1}$ and $k_{F2}$. Then, in the absence of
impurities, the electron classical orbit is also an ellipse with the main
radii
\begin{equation}
R_{c1}\approx 2\nu  /k_{F1}~,~R_{c2}\approx 2\nu /k_{F2}.
\label{Rci}
\end{equation}%
This equation is a particular (elliptic) case of the well-known
property that in magnetic field the electron orbits in coordinate
and momentum spaces differ by the 90 degrees rotation only. The
in-plane anisotropy of the localization length $\xi $ is the same
as the anisotropy of the cyclotron radius $R_{c}$, $\xi
_{i}\propto R_{c\,i}$, because the electron diffusion along the
conducting layer goes via the jumps from one cyclotron orbit to
another due to impurity scattering, and the average jump distance
is proportional to the size of the electron orbit in this
direction. Hence, the localization length in two main directions
is
\begin{equation}
\xi _{1}\approx R_{c\,1},~\xi _{2}\approx R_{c\,2}.  \label{xi2}
\end{equation}

Any localized electron state is covered by a polaron, which comes from the
static lattice deformation and from the rearrangement of the neighboring
electrons. Since the electron localization length $\xi \sim R_{c}$ is larger
than the Debye screening radius, the charge density $\rho \left( \mathbf{r}%
\right) $ induced by this polaron follows the square of the electron wave
function amplitude $\left\vert \Psi _{e}\left( \mathbf{r}\right) \right\vert
^{2}$. Hence, this polaron creates a positive charge distributed on the area
\begin{equation}
A_{0}\approx \pi \xi _{1}\xi _{2}.  \label{A}
\end{equation}%
Along the interlayer direction the induced charge also follows the electron
wave function $\left\vert \Psi _{e}\left( z\right) \right\vert ^{2}$\ and is
distributed within one crystal layer, because otherwise there would be a
nonuniform charge distribution along the $z$-axis. This polaron lowers the
energy of each localized electron.

\begin{figure}[tbh]
\includegraphics[width=0.49\textwidth]{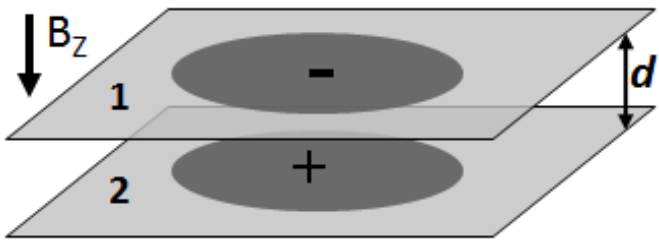}
\caption{The schematic view of the charge distributions (dark-gray areas) of
polaron (lower layer) and of the electron (upper layer) after an electron
jump to the adjacent layer. }
\label{Scheme}
\end{figure}

To jump to the adjacent conducting layer the electron now must overcome the
Coulomb attraction of polaron, which creates the potential barrier. To
estimate the value $E_{c}$ of this Coulomb barrier, consider the following
electrostatic problem, schematically shown in Fig. \ref{Scheme}. The flat
capacitor has charge $e$ separated by the interlayer distance $d$. The
energy stored in this capacitor is
\begin{equation}
E_{c}\approx 2\pi e^{2}d/\varepsilon A_{0},  \label{Ec}
\end{equation}%
where $\varepsilon $ is the dielectric constant of the media.

\section{Estimates of interlayer magnetoresistance at the metal-insulator
transition}

The Coulomb energy $E_{c}$, given by Eq. (\ref{Ec}), affects the
interlayer electron transport when it becomes comparable to the
band-width in the interlayer direction: $ E_{c}\gtrsim 4t_{z}$.
Substituting Eqs. (\ref{Rci})-(\ref{A}) into Eq. (\ref{Ec}), one
can estimate the crossover magnetic field $B_{c}$, above which
the Coulomb energy hampers the interlayer electron transport:%
\begin{equation}
B_{c}\approx F\sqrt{8\varepsilon t_{z}/e^{2}d\,k_{F1}k_{F2}},  \label{Bc}
\end{equation}%
where $F=k_{F1}k_{F2}c\hbar /2e$ is the MQO frequency. Sometimes,
it is convenient to use the crossover LL filling number $\nu
_{c}$, which corresponds to the field in Eq. (\ref{Bc}):
\begin{equation}
\nu _{c}\approx \sqrt{e^{2}dk_{F1}k_{F2}/8\varepsilon t_{z}},  \label{Nc}
\end{equation}

\begin{figure}[tbh]
\includegraphics[width=0.49\textwidth]{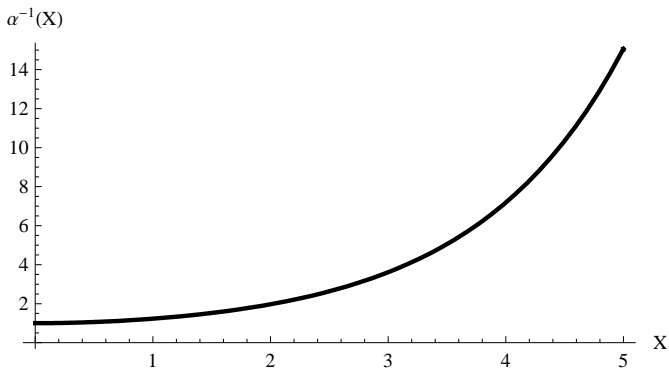}
\caption{The function $\protect\alpha ^{-1}\left( X\right) $ from Eq. (%
\protect\ref{al}). The predicted ratio $R_{zz}\left( B_{z}\right)
/R_{zz}\left( 0\right) \sim \protect\alpha ^{-1}\left( X\right) $
is given by the same curve, because $X\propto B_{z}$, as follows
from Eqs. (\protect \ref{X}),(\protect\ref{A}) and
(\protect\ref{xi2}).} \label{FigRzz}
\end{figure}

At $B>B_{c}$, or at $\nu <\nu _{c}$, the interlayer electron transport
becomes thermally activated. One could expect that at $B>B_{c}$\
\begin{equation}
\sigma _{zz}\left( B_{z}\right) /\sigma _{zz}\left( 0\right) \sim \exp
\left( -E_{c}/T\right) .  \label{alpha0}
\end{equation}%
This simple formula does not take into account two effects. First, the
finite interlayer transfer integral effectively reduces the activation
energy $E_{c}$ needed for an electron to jump to the next layer by the
bandwidth $4t_{z}$. This can be approximately taken into account by an
effective increase of temperature in Eq. (\ref{alpha0}) according to the rule%
\begin{equation}
T\rightarrow T^{\ast }\approx T+4t_{z}.  \label{T}
\end{equation}%
Second, the finite probability of interlayer jumping delocalizes
the electrons, which increases the effective in-plane electron
localization length $\xi $ and reduces the Coulomb energy $E_{c}$
in Eq. (\ref{Ec}). Third, the electron localization length vary
from one state to another, and one actually has a distribution of
localization length with function $P\left( \xi \right) $. This
function has maximum at $\xi \sim R_{c} $ and decrease rapidly at
$\xi \gg R_{c}$. However, the tails of the function $P\left( \xi
\right) $, which play the crucial role for the in-plane
conductivity, are also important for the interlayer transport
because of the exponential dependence on $\xi $ in Eq.
(\ref{alpha0}). It can be argued that the probability that the
electron localization area $A$ is much larger than $A_{0}\approx
\pi R_{c1}R_{c2}$ decreases exponentially:
\begin{equation}
P\left( A\right) \sim \left( A/A_{0}\right) \exp \left( -A/A_{0}\right) .
\label{PA}
\end{equation}%
Combining Eqs. (\ref{Ec}), (\ref{PA}) and (\ref{alpha0}) we obtain the
suppression factor of interlayer conductivity%
\begin{eqnarray}
\frac{\sigma _{zz}\left( B_{z}\right) }{\sigma _{zz}\left( 0\right) } &\sim
&\int_{0}^{\infty }\frac{dA}{A_{0}}\frac{A}{A_{0}}\exp \left( \frac{-A}{A_{0}%
}\right) \exp \left( \frac{-2\pi e^{2}d}{\varepsilon T^{\ast }A}\right)
\notag  \label{alpha} \\
&=&X^{2}K_{2}\left( X\right) /2\equiv \alpha \left( X\right) ,  \label{al}
\end{eqnarray}%
where $K_{2}\left( X\right) $ is the modified Bessel's function of the
second kind, and
\begin{equation}
X\equiv \sqrt{8\pi e^{2}d/\varepsilon T^{\ast }A_{0}}=\sqrt{4E_{c}/T^{\ast }}%
.  \label{X}
\end{equation}%
At $X\gg 1$ Eq. (\ref{al}) simplifies to
\begin{equation}
\sigma _{zz}\left( B_{z}\right) /\sigma _{zz}\left( 0\right) \sim \sqrt{\pi }%
\left( X/2\right) ^{3/2}\exp \left( -X\right) .  \label{alphaS}
\end{equation}%
In Fig. \ref{FigRzz} we plot the inverse function $R_{zz}\left( B_{z}\right)
/R_{zz}\left( 0\right) =\sigma _{zz}\left( 0\right) /\sigma _{zz}\left(
B_{z}\right) \propto \alpha ^{-1}\left( X\right) $.

\section{Comparison with experiments}

The increase of interlayer magnetoresistance described above is rather
general. It has been observed in many compounds, including heterostructures%
\cite{Kuraguchi2003} and various organic metals.\cite%
{WosnitzaPRL2001,Wosnitza2002,MarkPRL2006,Incoh2009,Zuo1999,Nam2001,ChCh}
This effect, probably, also occurs in pnictides and cuprate high-Tc
superconducting materials. For comparison with experiment we choose two
completely different materials: the GaAs/AlGaAs semiconductor superlattice%
\cite{Kuraguchi2003} and the organic metal $\beta $\textquotedblright
-(BEDT-TTF)$_{2}$SF$_{5}$CH$_{2}$CF$_{2}$SO$_{3}$.\cite{Zuo1999} This choice
is stipulated by the small value of the interlayer transfer integral $t_{z}$
in both these materials, which makes possible the in-plane electron
localization even at moderate magnetic field. In GaAs/AlGaAs superlattices
the interlayer transfer integral $t_{z}$ can be controlled artificially. In
the sample \#2 of GaAs/AlGaAs in Ref. \cite{Kuraguchi2003} $%
t_{z}=0.03meV=0.35K$ and the interlayer distance $d\approx 300\mathring{A}$.
In $\beta $\textquotedblright -(BEDT-TTF)$_{2}$SF$_{5}$CH$_{2}$CF$_{2}$SO$%
_{3}$ the small value of $t_{z}\lesssim 0.2K$ appears naturally because of
the large interlayer distance $d\approx 17.5\mathring{A}$.\cite{Wosnitza2002}

\subsection{GaAs/AlGaAs heterostructures}

The area electron concentration $n_{e}=k_{F}^{2}/2\pi =6\times
10^{11}cm^{-2} $ in the sample \#2 of GaAs/AlGaAs in Ref. \cite%
{Kuraguchi2003} corresponds to the Fermi momentum $k_{F}=2\times
10^{6}cm^{-1}$. This gives the cyclotron radius $R_{c}=\hbar
k_{F}c/eB_{z}=130nm/B_{z}\left[ T\right] $. Note that the magnetic length $%
l_{H}=\sqrt{\hbar c/eB_{z}}\approx 2.5\times 10^{-6}cm/\sqrt{B_{z}\left[ T%
\right] }$ is of the same order as $R_{c}$, \ and the filling
factor $\nu \sim 1$. The MQO frequency $F=k_{F}^{2}c\hbar
/2e\approx 11.7T$. The MQO and QHE are not observed because they
are smeared completely by high temperature $T=75K$. However, the
localization of the electron states is preserved and affects the
electron transport. Substituting $\varepsilon =12.4$ we obtain $
E_{c}\approx 2\pi e^{2}d/\varepsilon \pi R_{c}^{2} \approx
1.6K\times B_{z}^{2}\left[ T\right] $ and $X=\sqrt{4E_{c}/T^{\ast
}}\approx 0.3B_{z}\left[ T\right] $. Substituting this into Eq.
(\ref{al}) we obtain the dependence $R_{zz}\left( B_{z}\right) $.
This dependence $R_{zz}\left( B_{z}\right) $ is shown in Fig.
\ref{FigRzz}, where one only has to change the axis labels: on the
ordinate axis $\alpha ^{-1}\left( X\right) \rightarrow $
$R_{zz}\left( B_{z}\right) /R_{zz}\left( 0\right) $, and on the
horizontal axis $X\rightarrow 0.3B_{z}\left[ T\right] $. The
interval $0<B_{z}<13T$ of magnetic field in Fig.1d of Ref.
\cite{Kuraguchi2003} corresponds to the interval $0<X<4$ in Fig.
\ref{FigRzz}. This calculated dependence $R_{zz}\left(
B_{z}\right) $ agrees very well with the experimental data plotted
in Fig.1d of Ref. \cite{Kuraguchi2003} at $\theta =0^{\circ }$,
provided that there is a constant upward shift of the experimental
curve $R_{zz}\left( B_{z}\right) $ in Ref. \cite{Kuraguchi2003}
which, probably, came from in-series resistances in the
experimental setup. The crossover field, given by Eq. (\ref{Bc}),
$B_{c}\approx 0.5T$ also agrees with the experimental data in
Fig.1d of Ref. \cite{Kuraguchi2003}. Unfortunately, there is no
experimental data on the temperature dependence of the interlayer
magnetoresistance in Ref. \cite{Kuraguchi2003} to compare with out
theoretical prediction.

\subsection{Organic metal $\protect\beta $-(BEDT-TTF)$_{2} $SF$_{5}$CH$_{2}$%
CF$_{2}$SO$_{3}$}

\begin{figure}[tb]
\includegraphics[width=0.49\textwidth]{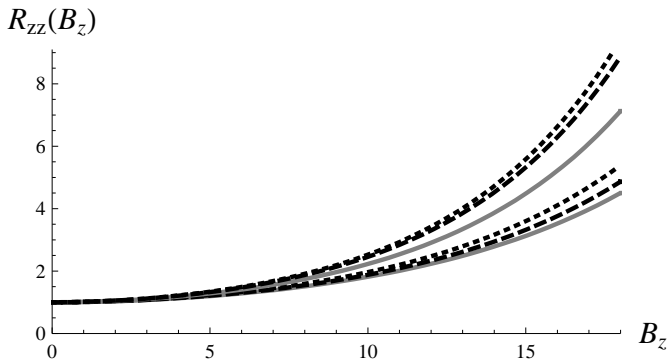}
\caption{The predicted field dependence of the background (non-oscillating) magnetoresistance $%
R_{zz}\left( B_{z}\right) /R_{zz}\left( 0\right) $ in $\protect\beta $%
-(BEDT-TTF)$_{2}$SF$_{5}$CH$_{2}$CF$_{2}$SO$_{3}$ at $t_{z} =0.2K$
for six values of temperature: $T=0.66$ K, $0.565$ K, $0.455$ K,
$0.218$ K, $80$ mK, and $47$
mK. The lower the temperature, the larger the magnetoresistance $%
R_{zz}\left( B_{z}\right) $, i.e. the highest curve in this plot corresponds
to the lowest temperature. The values of temperature are taken to be the
same as in Fig. 1 of Ref. \protect\cite{Zuo1999}.}
\label{FigRzzBeta}
\end{figure}

The in-plane Fermi surface in the organic compound $\beta $-(BEDT-TTF)$_{2}$%
SF$_{5}$CH$_{2}$CF$_{2}$SO$_{3}$ is an elongated ellipse with area\cite%
{FSBrooks} $\pi k_{F1}k_{F2}=0.0192\,\mathring{A}^{-2}$, which is about 5\%
of the first Brillouin zone. The interlayer distance $d\approx 17.5\mathring{%
A}$. Unfortunately, we are not aware of any measurements of the dielectric
constant $\varepsilon $ in $\beta $-(BEDT-TTF)$_{2}$SF$_{5}$CH$_{2}$CF$_{2}$%
SO$_{3}$. In organic metals the dielectric constant $\varepsilon $ varies in
the interval from $8$ to $50$ and may depend on magnetic field, pressure and
other external parameters.\cite{Dressel1994} Therefore, for the estimates
below, we take the average value $\varepsilon \approx 20$. The exact value
of the interlayer transfer integral $t_{z}$ in $\beta $-(BEDT-TTF)$_{2}$SF$%
_{5}$CH$_{2}$CF$_{2}$SO$_{3}$ is not known yet, but the experimental
estimates of the maximum value of $t_{z}$ give $t_{z}\leq 18.5\mu eV=0.21K$.
\cite{Wosnitza2002} Below we take $t_{z}=0.2K$. The other parameters are $%
d=17.5\,\mathring{A}$ and $F=200$ Tesla. Substituting this into Eqs. (\ref%
{Bc}) or (\ref{Nc}) we obtain the critical magnetic field $B_{c}\approx 8.5T$
and the LL filling number $\nu _{c}\approx 23$ in a qualitative agreement
with the experimental results in Refs. \cite{Zuo1999,WosnitzaPRL2001}.
Substituting the parameters to Eq. (\ref{Ec}) we obtain $E_{c}\approx 0.012%
\left[ K\right] \times B_{z}^{2}\left[ T\right] $ and $X=\sqrt{%
4E_{c}/T^{\ast }}\approx B_{z}\left[ T\right] \sqrt{0.05\left[
K\right] /\left( T+4t_{z}\right) }$. Substituting this into Eq.
(\ref{al}) we obtain the dependence $R_{zz}\left( B_{z},T\right) $
of the background
magnetoresistance, averaged over MQO. This predicted dependence $%
R_{zz}\left( B_{z}\right) /R_{zz}\left( 0\right) $ is shown in Fig. \ref%
{FigRzzBeta} for six different values of temperature: $T=0.66$ K, $0.565$ K,
$0.455$ K, $0.218$ K, $80$ mK, and $47$ mK. These values of temperature are
taken to be the same as in Fig. 1 of Ref. \cite{Zuo1999}, to simplify the
comparison with experiment. As in Fig. 1 of Ref. \cite{Zuo1999}, the highest
curve in Fig. \ref{FigRzzBeta} corresponds to the lowest temperature.
Comparison of Fig. \ref{FigRzzBeta} and Fig. 1 of Ref. \cite{Zuo1999} shows,
that the agreement of the above theory with experiment is very good, both in
the temperature and field dependence of magnetoresistance.

\section{Conclusion}

Above we developed a theory to describe the interlayer
magnetoresistance in the extremely anisotropic quasi-2D layered
metals. Our model takes into account the polaron effect. The
polarons dress the localized electron states, which prevents the
interlayer electron jumps by creating a Coulomb barrier. The
energy of this Coulomb barrier, given by Eqs. (\ref{Ec}) and
(\ref{A}), rapidly increases with the increase of the out-of-plane
component $B_{z}$ of magnetic field because of the decrease of the
electron localization length $ \xi \left( B_{z}\right) $, given by
Eqs. (\ref{xi1}) and (\ref{Rc}) [or by Eqs. (\ref{xi2}) and
(\ref{Rci}) in the case of in-plane anisotropy]. The obtained
temperature and field dependence of the interlayer conductivity is
given by Eqs. (\ref{al})-(\ref{alphaS}). The comparison with
experimental data on the field and temperature dependence of
interlayer magnetoresistance in
GaAs/AlGaAs heterostructures and in organic metal $\beta $-(BEDT-TTF)$_{2}$SF%
$_{5}$CH$_{2}$CF$_{2}$SO$_{3}$ is performed in Sec. IV. This
comparison shows the nice qualitative and quantitative agreement
of the proposed theory with experiments in very anisotropic
layered metals even without any fitting parameters. It also
explains long-standing problem of the strong field dependence of
interlayer magnetoresistance in such compounds. The proposed
theoretical model also allows to use the experiments on interlayer
magnetoresistance to investigate the electron states, localized by
magnetic field and disorder.

\bigskip

The work was supported by Russian Foundation for Basic Research, Foundation
"Dynasty" and by LEA ENS-Landau exchange program.

\end{document}